\documentclass[12pt]{article}
\def\ML{\mbox{$M/L$}}
\usepackage{sc3conf}
\usepackage{amsfonts}
\usepackage{epsfig}

\begin{document}

\title{Mass-to-light ratios in early-type galaxies and the dark matter content}

\authors{\normalsize M. Capaccioli,\adref{1,2}
 N.R. Napolitano,\adref{1} 
 and M. Arnaboldi\adref{1,3}}

\addresses{\1ad INAF--Capodimonte Astronomical Obs., via Moiariello 16,
I-80131 Naples, Italy\\
 \2ad Dept. of Physical Sciences, University ``Federico II'', Naples, Italy\\
 \3ad INAF--Pino Torinese Astronomical Obs., via Osservatorio 20, I-10025 Pino Torinese}

\maketitle
\begin{abstract} 
The distribution of the radial trends of the mass-to-light ratios (\ML) within an assorted sample of early-type galaxies is discussed. 
Three classes of galaxies are identified according to their \ML\ gradients.
Two such classes are characterized by the presence or by the absence of a radial gradient of the dark-matter (DM) distribution.
A third class contains objects which are likely undergoing interaction; they exhibit steep \ML\ gradients which are possibly the result of a wrong assumption on their equilibrium conditions.
Finally, a possible correlation between DM content and morphological types is briefly discussed.
\end{abstract}
\section{Introduction}
The currently favored hierarchical theories of galaxy formation predict that galaxies should be surrounded by DM halos with a rather universal profile (NFW, \cite{33,34}). 
In this scenario elliptical galaxies are mostly formed through merging events and their angular momenta largely reside in the outer halos \cite{51,3,49}.
From the observational point of view, while a number of studies based on various mass tracers (inner stellar kinematics, X-ray emission from hot gas, and gravitational lensing) seem actually to favor the occurrence of a significant amount of DM in the halos (\cite{18,27,23} and references therein), there is only marginal evidence for some DM inside $\sim2R_e$\footnote{The effective radius $R_e$ characterizes the isophote encircling half of the total light of the galaxy} \cite{38,42}.\\
The presence, and in particular the ubiquity, of DM in early-type galaxies (E and S0's) is nowadays a debated topic. 
Saglia et al. \cite{41} first suggested the possible existence of at least two classes of early-type galaxies: a first class with decreasing velocity dispersion where no DM is required, and a second one with constant or rising velocity dispersion, consisting of the most likely DM candidate systems. 
This picture appears in contrast with the predictions from the hierarchical galaxy formation theory. 
Unfortunately, the studies based on the stellar kinematics from integrated-light spectra are confined by technical reasons to inside the inner luminous regions ($R\sim R_e$). 
Untill recently observations at larger radii were made difficult by the lack of suitable and easily interpreted tracers as they are, for instance, the HI rotation curves for spiral galaxies. 
A sampling of the outer kinematics of early-type galaxies might better constrain both angular momentum and DM distribution in these systems and test the predictions of the galaxy formation theories, in particular concerning the ubiquity of DM. 
The need to extend outwards the kinematical mapping has motivated, in the last decades, the use of new classes of tracers working at large distances from galaxy centers.
Besides the already mentioned X-rays from hot gas and gravitational lensing, globular clusters (GCs) represent a further promising class of dynamical tracers \cite{19,25,11}. 
None of these tracers, however, is really limitation-free. 
Hot gas is highly collisional and sensitive to both inner and outer energetic mechanisms (inner radio jets, galaxy-galaxy interactions, etc.) which might disturb its equilibrium. 
GCs possibly belong to a population decoupled from that of the general stellar content and have experienced a different evolution: they can be suitably used as dynamical tracers, but their kinematics, and in particular their rotational structure, can be different and give biased angular momentum estimates. 
Gravitational lensing is a really promising technique, but it is inappropriate for systematic studies, being based on random events.\\ 
A novel powerful class of kinematical tracers has been recognized in the planetary nebulae (PNe). These are Post-AGB stars in their final evolutionary phase, when they are expelling the outer shells of their atmosphere disclosing the inner hot star (${\sim10^{5}}~$K). 
This\ ionizes the planetary envelope thus producing, in particular, the typical powerful [OIII] emissions at $\lambda 5007 \AA$ and $4959 \AA$. 
By these spectral features, the PNe are easily detectable, in principle in much larger numbers than GCs, and their radial velocities measurable at large distances ($D>15$ Mpc).
PNe have already been used successfully to probe the kinematics of the outer regions of early-type galaxies \cite{36,10,48,22,1,2,14,32}. 
In these systems they match the light distribution \cite{9,16} and the kinematical behavior of the stars where overlapping with the integrated stellar-light data \cite{2,22,32}. 
Thus PNe are the natural candidates to gauge the stellar kinematics in the outskirts of ellipticals.
One main limitation of the technique has been so far the small size of the PNe samples: typically, about 50 PNe radial velocities with 4-m class telescopes and multi-objects spectroscopy\footnote{But for the nearby giant galaxy NGC 5128 where 433 PNe radial velocities were collected \cite{22}}. 
Although Napolitano et al. \cite{30} have proved the statistical reliability of small samples, larger samples are needed to improve the precision of the estimates, and deeper samples are required to apply the technique to more distant galaxies.
To this end a new generation dedicated spectrograph, the Planetary Nebulae Spectrograph (PN.S), has been recently built and commissioned by an international consortium (see http://www.astro.rug.nl/~pns/).\\
In this paper we compare the \ML\ measurements in the halo regions of early-type galaxies from the published papers based on the PNe kinematics, with the estimates in the inner luminous parts. 
The \ML\ values at large radii range from $\sim~2R_e$ to $6R_e$, and concern a very heterogeneous sample of early-type galaxies: a total of 9 systems, hereafter referred to as sample A, which will allow us to draw a picture of the actual understanding on the DM distribution in ellipticals out to unprecedented galactocentric distances. 
In order to improve the statistics, we added to our catalogue 5 galaxies (hereafter sample B) for which the stellar kinematical mapping has the largest radial extension ($R\ge 2 R_e$). 
\section{The catalogue and the \ML\ estimates}
Table 1 reports, for each galaxy from our two samples, the distance and the inner (stellar) and outermost mass-to-light ratio estimates ($M/L_{B,*}$ and $M/L_{B,h}$ respectively) from the quoted references, expressed in solar units. 
Data for the sample B have been mostly derived from \cite{28} adopting $h_0=0.85$, a value similar to that derived from the PNe luminosity function on which the large part of the distances for the sample A are based. 
Gerhard et al. \cite{18} derived $M/L_B$ values for these same galaxies adopting $h_0=0.65$; we have checked and confirmed consistency between the two sets, once scaled to the same distance.
The \ML\ estimates of Table 1 have been obtained through different dynamical analysis techniques. 
In particular, the PNe values are mostly based on the inversion of Jeans equations with simplified assumptions on the anisotropy parameter. 
{\em A posteriori}, this appears as a reasonable approach in view of the substantial consistency obtained with the results from the more sophisticated techniques used for the inner stellar kinematics (see \cite{32}, for instance). 
The large error on the PN estimates comes essentially from the limited kinematical data rather than from the robustness of the approach. 
When the literature lack indications on the \ML\ errors, we assumed a 10\% error in the inner regions and a 30\% error at large radii, which are very conservative error budgets for such studies.\\
\begin{table}[t]
\vspace{-1.5cm}
\caption{\footnotesize Data catalogue for the early-type galaxies considered in this paper. 
Sample A includes objects with available PNe kinematics, sample B with an extended integrated stellar-light kinematical coverage ($R \ge 2 R_e$).}
\bigskip
\flushleft
\scriptsize
\centering
\vspace{-1cm}
\begin{tabular}{lccccc}
\hline \hline
 \multicolumn {5}{c}{\normalsize{\bf Galaxy catalogue}}\\
\hline \hline 
\noalign{\smallskip}
Galaxy & $D/$Mpc & $M/L_{B,*}$ ($R/R_e$) & $M/L_{B,h}$ ($R/R_e$)& Ref \\
\hline
\multicolumn {5}{c}{Sample A}\\
\hline
 M32 & 0.8 & 2.5 (0.25) & 6 (6)$^a$ & Magorrian \& Ballantine (2001)\\
 NGC1316 & 16.9 & 4.3 (0.5) & 8 (2)& Arnaboldi et al. (1996)\\
 NGC1399 & 17 & 12 (1) & 20 (5.4)-26 (8.8) & Bicknell et al.(1989),Napolitano et al. (2002) \\
 NGC3379 & 10 & 7 (0.5) & 7 (2.2) & Ciardullo et al. (1993)$^b$ \\
 NGC3384 & 10.1 & 5 (1) & 9 (7) & Busarello et al.(1996), Tremblay et al. (1995) \\
 NGC4406 & 17.8 & 12 (0.2) & 13 (1.2) & Arnaboldi et al. (1996) \\
 NGC4472 & 20 & 3.6 (0.3) & 8 (2.6)& Romanowsky et al. (2001) \\
 NGC4697 & 10.5 & 11 (0.5) & 11 (3)& M\'endez et al. (2001)$^c$ \\
 NGC5128 & 3.5 & 3.9 (0.3) & 10 (4.8) & Hui et al. (1995) \\
\hline
\multicolumn {5}{c}{Sample B}\\
\hline
 NGC1379 & 17.9 & 3 (0.4) & 3 (2.5) & Magorrian\&Ballantine (2001) \\
 NGC1700 & 34 & 4 (0) & 8 (4.6) & Statler et al. (1999) \\
 NGC2434 & 20 & 8 (0.5) & 20 (4) & Magorrian\&Ballantine (2001) \\
 NGC4464 & 15.3 & 10 (0.5) &12 (4) & Magorrian\&Ballantine (2001) \\
 NGC5846 & 26 & 9 (0.5) & 20 (3.2) & Magorrian\&Ballantine (2001) \\
\hline\hline
\multicolumn {5}{l}{}\\
\multicolumn {5}{l}{$^a$ Nolthenius \& Ford (1986): constant $M/L_B$=2-4 for a $D=0.67$ Mpc.}\\
\multicolumn {5}{l}{$^b$ Gebhardt et al. (2000): constant $M/L_V$=4.85; Magorrian \& Ballantine: $M/L_B=11$ at $R=6R_e$.}\\
\multicolumn {5}{l}{$^c$ Napolitano 2001 (PhD thesis): $M/L_B=12.4\pm1.0$ accounting for rotation and $D=10.5$ Mpc.}\\
\noalign{\smallskip}
\end{tabular}
\label{omega}
\end{table}
~\\
\subsection{Some remarks on individual galaxies}
\underline{\em NGC3379-NGC3384} --
Schneider \cite{45} studied the kinematics of the 200 kpc diameter HI ring around these two galaxies measuring an enclosed total mass of $5.6\times10^{11}M_{\odot}$. 
Tremblay et al. \cite{48} discussed the possibility that these two objects are surrounded by a very extended dark halo (out to 45 kpc from NGC3384 center) with a mass which should be one third of the total mass of the two galaxies as evaluated via PNe kinematics to be in agreement with the \cite{45}
estimate.
Based on photometric studies, Capaccioli et al. \cite{8} suggested that NGC3379, rather than a standard elliptical, could be a misclassified S0 seen close to face--on. 
This view has been supported by recent kinematical studies from \cite{47}.\\
\underline{\em NGC1316, NGC1700, NGC4406, NGC4472, NGC5128, NGC5846} -- These are giant/normal ellipticals which are recognized to be major merger product candidates.\\
\underline{\em NGC1399} -- 
It is the cD galaxy dominating the Fornax cluster. 
Napolitano et al. \cite{32} have shown that kinematic substructures in the PNe radial velocity field are compatible with a fly-by encounter with the nearby companion NGC 1404; i.e. NGC1399 outer regions are out of equilibrium. 
This is possibly supported by the very irregular density distribution of the X-ray emissivity (see \cite{37}) which could be expected in such an interaction scenario. 
Interaction with NGC1404 is also suggested from GCs specific density studies \cite{25}. 
In Table 1 we report the $M/L_B$ value obtained under this encounter scenario, while the standard equilibrium estimates obtained from \cite{32} will be discussed later in the text.\\
\underline{\em NGC4697} -- 
The PNe system shows a decreasing rotations out of $1R_e$. 
M\'endez et al. \cite{29} quoted a constant $M/L_B$=11 in agreement with \cite{6} when scaled at the same distance. 
Napolitano \cite{31} also found $M/L_B=12.4\pm 1.0$ accounting for the rotation in the Jeans equations. However the distance of 10.5 Mpc for NGC4697 assumed in these studies is possibly too small for a Virgo member.
Assuming a more realistic distance $D=15$ Mpc, we obtain $M/L_B\sim 8$ in both cases.\\
\underline{\em NGC1379} -- 
E3 galaxy, which D'Onofrio et al. \cite{13} argued to be an S0 galaxy seen quite face--on.\\ 
\underline{\em NGC4464} -- 
Classified as E3, it was suggested to be possibly a spiral galaxy by \cite{26}, while in \cite{20} evidence is found that it could be a double component system, with an inner fast rotating disk extending to $5'' \sim 0.4$ kpc and an outer slowly rotating bulge. 
This rotational structure looks similar to that of NGC4697 and also to that of NGC3379 as recently derived from a preliminary analysis of the data obtained by the PN.S team during the commissioning of this spectrograph.\\ 
\underline{\em M32} -- 
E3 galaxy with evidence of tidal interaction \cite{36,50} or possibly a merger remnant \cite{12}.\\ 
\underline{\em NGC2434 and NGC5846} -- 
These galaxies have been recognized as two possible systems undergoing interaction. 
In \cite{15,44} a close interaction between NGC2434 and NGC2442 is invoked to explain the visual structures of this latter galaxy. 
This scenario has been recently confirmed in \cite{40}. 
Higdon et al. \cite{21} modeled a fly-by encounter between NGC5846 and NGC5850 as the most favorable explanation for the irregularity and asymmetries in both the morphology and the kinematics from the HI emission.
\section{Discussion of the collected results}
The $M/L_B$ values for the 14 galaxies in our catalogue are plotted in Fig. 1. 
Here the inner $M/L_{B,*}$ and outermost $M/L_{B,h}$ values have been connected by straight lines to provide a visual check on their variation from the innermost to the outermost sampling radii.
In other words these lines do not account for the actual trend of the mass distribution. 
The dashed line and the long-short dashed line reproduce the results from equilibrium and tidal interaction models respectively, derived in \cite{32} for NGC1399. 
We point out the large spread in the $M/L_{B,*}$ estimates which probably comes mostly from the uncertainties on the assumed distances. 
In order to better homogenize the results and have a visual perception of the trend of the \ML\ ratios from the inner to the outer radii, in the right panel of Fig. 1 we shifted all the \ML\ pairs of values by the quantities $(-M/L_{B,*}+6)$ needed to obtain an arbitrary constant value $M/L_{B,*}=6$. 
Here we also plot with different symbols the expected shape for NFW \cite{34} density profiles with the scale factor $r_s$ equal to $1R_e$, $5R_e$, and an unrealistic $r_s=0$, all arbitrarily scaled in the vertical direction.\\
We identify a subset of galaxies with a clear constant $M/L_{B}$: NGC3379, NGC1379, NGC4664, and NGC4697. 
The large error bars for the $M/L_{B}$ values of M32 and NGC3384 do not make it clear whether these objects can be considered constant $M/L_{B}$ galaxies or not. 
However, the $M/L_{B}$ values for all these galaxies (including NGC4697, if distance corrected as discussed in the previous Section) look in agreement with the typical stellar values for evolved populations. 
This does not make it compulsory to assume a uniform overall mass distribution from a diffuse DM halo, which however could not be excluded (e.g., \cite{43} for analogy with spirals). 
In this sense we will refer to such a constant-$M/L_{B}$ galaxies as no-DM systems and conservatively include in the group, at this stage, also M32 and NGC3384. 
\begin{figure}[t]
 \begin{center}
 \vspace{-2.2cm}
 \hspace{-2.3cm}
 \epsfig{file=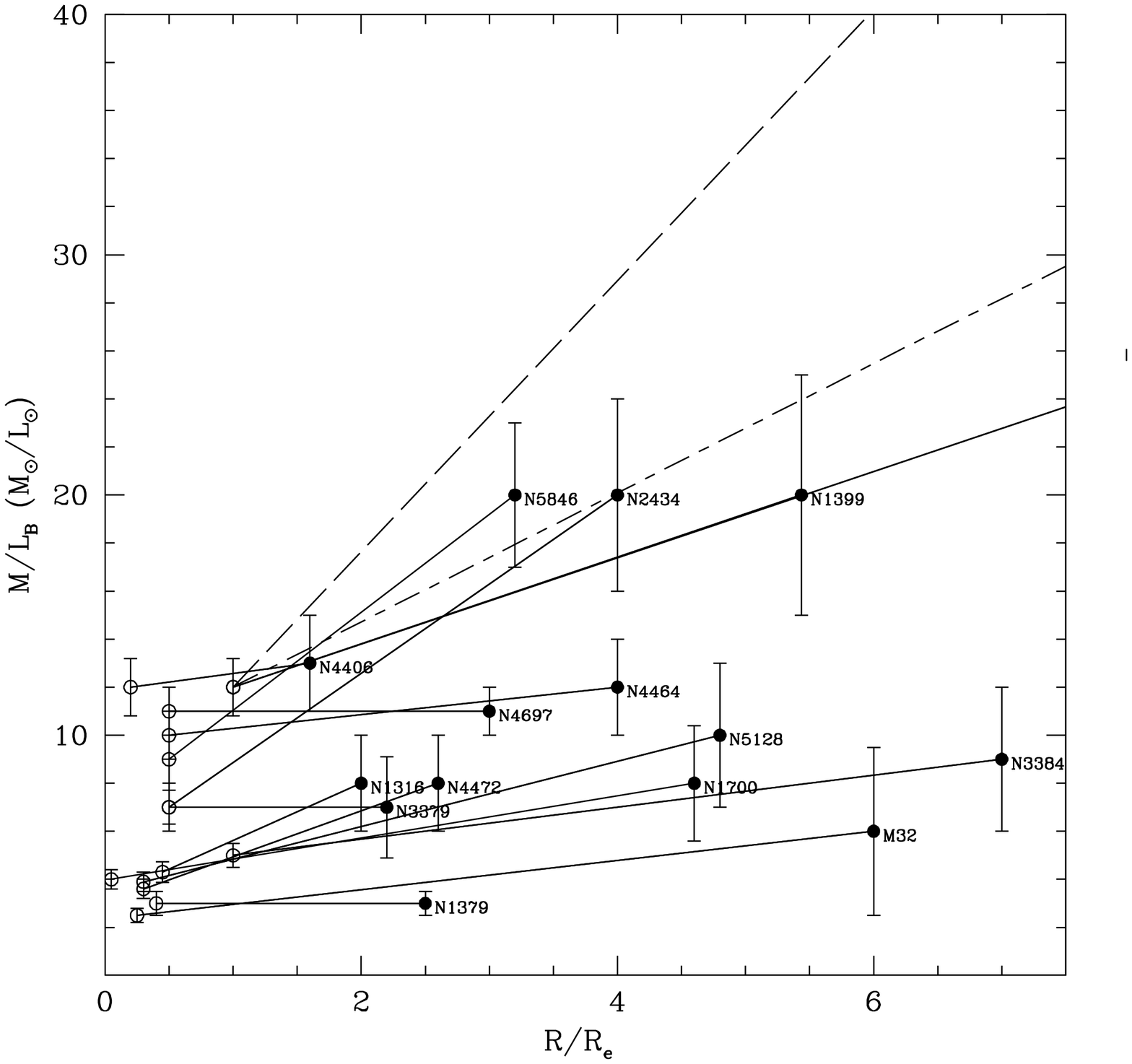,width=7.8cm,height=7.8cm}
 \epsfig{file=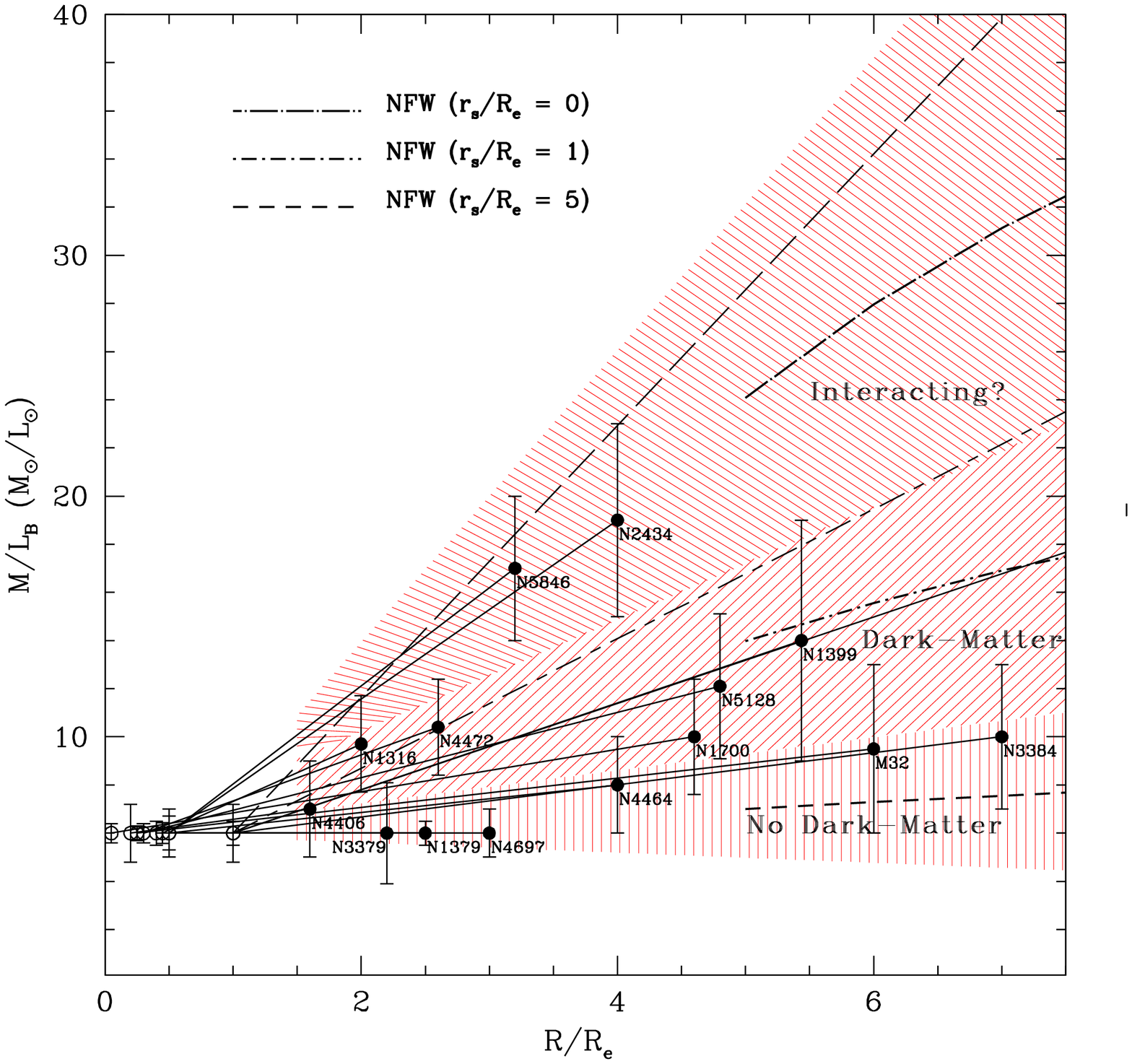,width=7.8cm,height=7.8cm}
 \caption{\small Mass-to-light ratios for the galaxy catalogue as in Table 1. Open points are $M/L_{B,*}$ estimates, full points are $M/L_{B,h}$s. Long dashed and lon--short dashed lines are the \ML\ model for NGC1399 considering simple equilibrium and the presence of a tidal radius respectively. NFW profiles are also plotted with different symbols, see discussion in the text.}\label{fig:1}
 \end{center}
\end{figure}
Increasing \ML\ ratios are obtained for the residual sample of galaxies, strongly advocating dark matter. 
All these galaxies except NGC2434 are classified as merger remnants (see previous Section). As a cD galaxy, NGC1399 has obviously experienced merging. 
Interestingly, the steepest gradients in the \ML\ ratios are shown by those galaxies which are thought to be interacting systems (NGC2434, NGC5846, and NGC1399, if equilibrium is assumed). Following Napolitano et al. \cite{32}, interactions could cause an overestimate of the mass, and thus of the \ML\ ratios, if standard equilibrium analysis is adopted (see dashed line for the equilibrium model of NGC1399 in Fig. 1). 
The trend in the growth of the $M/L_B$ for NGC5846, NGC2434, and NGC1399-equilibrium is more similar to the unrealistic NFW $r_s=0$ profile with respect to the case of $r_s=1R_e$. 
With this in mind we attempt a sub-classification of the DM systems: normal systems and interacting candidates. 
Again, the large error bars do not allow a strict separation between these classes to be made, but we believe that the possibility of considering biases in the actual mass distributions of elliptical interacting candidates to be a more conservative way to approach the understanding of their inner structure.\\
If we exclude the interacting candidates, we notice a singular coincidence. 
While the no-DM galaxies are all classified as {\em disky} or flattened or S0s misclassified as ellipticals, the normal-DM sample is built of post-merger systems, mostly {\em boxy}\footnote{The possibility that M32 is a merger remnants could put this galaxy in this second class. However, M32 classification is the most ambiguous.}. 
Could the DM content in ellipticals correlate with their morphology, and why?
A full interpretation of such a correlation, if real, is out of the scope of this paper. 
Here, we want to introduce some hints for further discussion. \\
Bender \cite{4} and Nieto \cite{35} suggested a correlation of the photometrical properties with the evolution of early-types: {\em disky} ellipticals are galaxies which retain their original structure while {\em boxy} ones are more likely merger products. 
This is possibly supported by recent studies on the specific density of GCs in ellipticals \cite{24}. 
In this picture, an intriguing possibility, which arises from the results discussed here, is that the {\em boxy}-merger-products have a significant DM content, i.e. they are surrounded (more or less) by compact halos, while the {\em disky}-unperturbed systems (which have experienced a substantial passive evolution) have no DM or, in a more conservative way, reside in a very diffuse dark halo. 
How could this be reconciled with the galaxy formation theories?
More conservatively, a further explanation for the absence of DM around galaxies is that tidal interactions could have removed their dark halos. 
Eventually, these galaxies have evolved into rotating, mildly non-spherical systems, with a roughly isotropic velocity ellipsoid \cite{36,29}.
\section{Conclusions}
We have presented a collection of $M/L_B$ estimates found in the literature, making use of the inner stellar-light integrated data and the outer halo kinematics from PNe. 
Depending on the gradient of the $M/L_B$(R), we have roughly identified three classes of galaxies:
$a)$ no-DM objects with a constant $M/L_B$; 
$b)$ DM galaxies with an increasing $M/L_B$(R); and 
$c)$ interacting candidates systems with very steep $M/L_B$(R).
The mass distribution for this last class derived assuming equilibrium are possibly biased and cannot be considered representative of ellipticals.\\
These results seem to confirm the supposition by Saglia et al. \cite{41} of two classes of early-type galaxies: one with decreasing velocity dispersion and no-DM, and another with a constant/increasing velocity dispersion and significant DM.
Since hierarchical theories predict that ellipticals should be surrounded by DM halos, this oncoming evidence of no-DM systems needs to be reconciled within galaxy formation theories.
Finally, we have shown a possible correlation between classes $a$ and $b$ and the morphological classification: no-DM systems are preferentially {\em disky} while DM systems are mostly {\em boxy}-merger remnants.\\ \\
{\em Acknowledgements.} This work has been supported by funds granted by the Italian Ministry of University and Research (MIUR), research program year 2000 ref. prot.MM02918885.

\end{document}